\documentclass{elsart}
\usepackage{graphicx}
\usepackage{amssymb}

\begin{document}

\begin{frontmatter}

\title{Echo spectroscopy and Atom Optics Billiards}
\author{M. F. Andersen, A. Kaplan, T. Gr\"{u}nzweig and N. Davidson.}
\address{Department of Physics of Complex Systems, Weizmann Institute of Science, Rehovot 76100, Israel}
\begin{abstract}
We discuss a recently demonstrated type of microwave spectroscopy
of trapped ultra-cold atoms known as "echo spectroscopy" [M.F.
Andersen et. al., Phys. Rev. Lett., in press (2002)]. Echo
spectroscopy can serve as an extremely sensitive experimental tool
for investigating quantum dynamics of trapped atoms even when a
large number of states are thermally populated. We show numerical
results for the stability of eigenstates of an atom-optics
billiard of the Bunimovich type, and discuss its behavior under
different types of perturbations. Finally, we propose to use
special geometrical constructions to make a dephasing free dipole
trap.

\end{abstract}

\end{frontmatter}

\section{Introduction}

The subject of decoherence of a quantum system has witnessed a
renewed interest in the last years. In addition to being central
to our understanding of the transition from quantum to classical
physics\cite{Paz99}, it is of outmost importance in the blooming
field of Quantum Information\cite{Nielsen01}, where decoherence or
dephasing represent loss of information.

A quantum superposition state "collapses" due to the dissipative
interaction with the environment (e.g. in the case of trapped
atoms, the coupling to the electro-magnetic vacuum leads to
spontaneous scattering of photons). This interaction is
irreversible in nature. On the other hand, the macroscopic
(ensemble averaged) response of a quantum system prepared in a
superposition state decays due to the dephasing resulting from
local variations in the evolution of the system (e.g.
inhomogeneous broadening). By stimulating an effective ''time
reversal'', revivals of coherence in dephased systems (at least
partial ones) have been reported for spin echoes \cite{Hahn50} and
photon echoes \cite{Kurnit64,Allen87}, and more recently for a
motional wave packet echo using ultra cold atoms in a
one-dimensional optical lattice \cite{Buchkremer00}.

In this paper we discuss the technique of "echo spectroscopy" on
ultra cold atoms trapped in an optical dipole trap, which utilizes
a micro wave (MW) analog to the photon echo \cite{Andersen02b}. We
show that this technique can be used to study the quantum dynamics
of trapped atoms, even when many states are thermally populated.
We show that the stability of eigenstates under perturbations of
the Hamiltonian plays an important role for the echo signal.
Numerical calculations of this quantity for an atom-optics
Bunimovich stadium are presented. Next, by overcoming the
inhomogeneous shift, which is the main limiting factor for the
coherence time in dipole traps \cite{Kaplan02b}, echo spectroscopy
opens the way for using dipole trapped atoms in quantum
information and possibly for high precision spectroscopy. Finally,
our analysis leads to a proposal of a dephasing free dipole trap.

\section{Principles of Echo Spectroscopy}

One of the most widely used MW spectroscopy techniques is the so
called Ramsey method \cite{Ramsey56}. It works on two level atoms
after the following principle: The atoms are prepared in one of
their two states (denoted $\left| 1\right\rangle $). Next, a short
MW-pulse close to resonance transfers the atoms into a coherent
superposition state of the two levels (This pulse is usually
called a $\frac{\pi }{2}$  pulse), which is left to evolve for a
certain time, without MW-irradiation. Due to the detuning between
the atomic resonance frequency and the MW-frequency, the time
evolution will cause an atomic phase to develop in the rotating
frame of the MW field. A second $\frac{\pi }{2}$-pulse maps this
phase onto populations of the two levels, which can be detected.
If the MW-frequency is scanned, this phase difference is scanned,
and sinusoidal oscillations in the populations, known as Ramsey
fringes, are seen. When Ramsey spectroscopy is performed on an
ensemble of inhomogeneously broadened two level atoms, each atom
will acquire a slightly different phase in the dark period,
causing the macroscopic fringe contrast to decay when increasing
the time between the $\frac{\pi }{2}$-pulses. Thus the Ramsey
fringe contrast is a measure of the dehpasing of the system.

\begin{figure}[t]
\begin{center}
\includegraphics[width=3in]{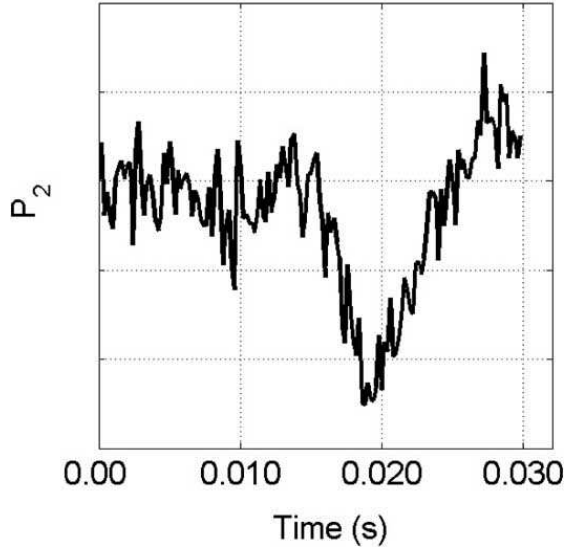}
\end{center}
\caption{Echo signal ($P_{2}$) measured as a function of time
between $\protect\pi $-pulse
and second $\protect\pi/2 $-pulse ($\protect\tau_{2}$) for fixed $\protect%
\tau_{1}$. A dip in $P_{2}$ is seen, showing a "coherence echo".
$\protect\tau_{2}=\protect\tau_{1}$} \label{fi1}
\end{figure}

The dephasing can be reversed by stimulating a {\it coherence
echo}. This is done by adding a population inverting pulse (a $\pi
$-pulse) between the two $\frac{\pi }{2}$-pulses. If the $\pi
$-pulse is added exactly in the middle between the two $\frac{\pi
}{2}$-pulses each part of the superposition state generated by the
first $\frac{\pi }{2}$-pulse will have spent an equal amount of
time in both levels at the time of the second $ \frac{\pi
}{2}$-pulse, and will therefore be exactly in phase with each
other. This means that a coherence echo have appeared at the time
of the second $\frac{\pi }{2}$-pulse even for a system that have
dephased completely before the $\pi $-pulse. When measuring the
populations the coherence echo is therefore observed by seeing
that all the atoms have
returned to the initial state after the $\frac{\pi }{2}$-$\pi $-$\frac{\pi }{%
2}$-pulse sequence. If the time ($\tau _{1}$) between the first $\frac{\pi }{%
2}$ and the $\pi $-pulse is kept constant, and the time between
the $\pi $ and the second $\frac{\pi }{2}$-pulse ($\tau _{2}$) is
swept, the coherence echo is seen as a decrease in the population
of $\left| 2\right\rangle $ for $\tau _{2}=\tau _{1}$ (see Fig.
\ref{fi1}). The population $P_{2}$ for $\tau _{2}=\tau _{1}$ is
referred in what follows as the "echo signal", where $P_{2}=0$
indicates full revival of coherence and $P_{2}=\frac{1}{2}$
indicates complete dephasing.

Experimental data shown in this paper is taken using two hyperfine levels of
the ground state of $^{85}Rb$ atoms trapped in a dipole trap. These two
levels ($\left| 5S_{1/2},F=2,m_{F}=0\right\rangle $, denoted $\left|
1\right\rangle $, and $\left| 5S_{1/2},F=3,m_{F}=0\right\rangle $, denoted $%
\left| 2\right\rangle $) are separated by the energy difference
$E_{HF}=\hbar \omega _{HF}$ where $\omega _{HF}=2\pi \times 3.036$
GHz is the hyperfine splitting. For spectroscopy of trapped atoms
we must consider the entire Hamiltonian including both the
internal and the external degrees of freedom of the atom. Since
the dipole potential is inversely proportional to the trap laser
detuning $\delta $ \cite{Cohen-Tannoudji92} there is a slightly
different potential for atoms in different hyperfine
states\footnote{For these two hyperfine levels the matrix elements
for the dipole interaction are identical, hence the relative
difference between the
two optical potentials is $\left( V_{1}-V_{2}\right) /V_{2}\simeq\frac{%
\omega_{HF}}{\delta}\sim10^{-3}$ for our experimental parameters.}
 and the Hamiltonian of our trapped two level atom can be written
as:
\begin{eqnarray}
H&=&H_{1}\left| 1\right\rangle \left\langle 1\right| +H_{2}\left|
2\right\rangle \left\langle 2\right| \nonumber \\
 &=&\left( \frac{p^{2}}{2m}+V_{1}\left( {\bf x}\right) \right) \left|
1\right\rangle \left\langle 1\right| +\left(
\frac{p^{2}}{2m}+V_{2}\left( {\bf x}\right) +E_{HF}\right) \left|
2\right\rangle \left\langle 2\right| ,  \label{hami}
\end{eqnarray}
where $p$ is the atomic center of mass momentum and $V_{1}$
[$V_{2}$] the external potential for an atom in state $\left|
1\right\rangle $ [$\left| 2\right\rangle $], much smaller than
$E_{HF}$. The atoms are initially prepared in their internal
ground state $\left| 1\right\rangle $. Their total wave function
can be written as $\Psi =\left| 1\right\rangle \otimes $ $\psi $,
where $\psi $ represents the motional (external degree of freedom)
part of their wave function. If a MW field close to resonance with
$\omega _{HF}$ is applied, transitions between the eigenstates of
the Hamiltonian corresponding to different internal states, can be
driven. The small size of optical traps (typically on the order of
$100$ $\mu m$ or less) as compared to the MW wavelength ($\sim $10
cm), ensures that the momentum of the MW photon can be neglected
(Lamb-Dicke regime \cite{Cohen-Tannoudji92}). Even though the
internal states of the atoms are described as a two-level system
Eq.\ref{hami} describes a multi-level system in two energy
manifolds. When the MW-field is added to the Hamiltonian in
Eq.\ref{hami} and the time-dependent Schr\"{o}dinger equation is
formulated it is seen that the transition matrix elements between
eigenstates of Eq. \ref{hami} are given by $C_{nn^{\prime
}}=\left\langle n^{\prime }\mid n\right\rangle \times
M_{1\rightarrow 2}$ where $M_{1\rightarrow 2}$ is the
free space matrix element for the internal state transition, and $%
\left\langle n^{\prime }\mid n\right\rangle $ is the overlap
between the initial motional eigenstate of $H_{1}$ and an
eigenstate of $H_{2}$. Therefore, when a strong and short MW pulse
is applied, the
motional part of the initial wave function is simply projected into $%
V_{2}\left( {\bf x}\right) $. If $V_{1}\left( {\bf x}\right) =V_{2}\left(
{\bf x}\right) $ then clearly $\left\langle n^{\prime }\mid n\right\rangle
=\delta _{nn^{\prime }}$. For a small enough potential difference
(perturbation) $\delta V=V_{2}\left( {\bf x}\right) -V_{1}\left( {\bf x}%
\right) $ the approximation $\left\langle n\mid n^{\prime }\right\rangle
\simeq \delta _{nn^{\prime }}$ is valid. In the general case, a projected
eigenstate of $H_{1}$ will not be an eigenstate of $H_{2}$, and therefore
will evolve in the new potential, causing the overlap $\left| \left\langle
n\left( t=0\right) \mid n\left( t=\tau \right) \right\rangle \right| $ to
decay ($\left| n\left( t=\tau \right) \right\rangle \equiv \exp \left( -i%
\frac{H_{2}}{\hbar }\tau \right) $ $\left| n\right\rangle $).

If Ramsey spectroscopy is performed on atoms initially prepared in an
eigenstate of $V_{1}$ characterized by the quantum number $n$, then the
probability to be in the internal state $\left| 2\right\rangle $ after the $%
\pi /2$-$\pi /2$ pulse sequence can be shown to be:
\begin{equation}
P_{2}=\frac{1}{2}\left( 1+\left\langle n\left| e^{-i\left( \frac{H_{2}}{%
\hbar }-\omega _{MW}+\Delta _{n}\right) \tau }\right| n\right\rangle \cos
\left( \Delta _{n}\tau \right) \right) ,
\end{equation}
where $\omega _{MW}$ is the MW-frequency, $\tau $ is the time between the
pulses, and $\Delta _{n}$ is a generalized, state-dependent, detuning (it
reduces to the detuning when only two levels are coupled by the MW-field), defined so $\left\langle n\left| \exp \left( -i\left( \frac{%
H_{2}}{\hbar }-\omega _{MW}+\Delta _{n}\right) \tau \right) \right|
n\right\rangle $ is real and positive. Scanning $\omega _{MW}$ for a fixed $%
\tau $ yields the usual Ramsey fringes with a contrast given by $%
\left\langle n\left| \exp \left( -i\left( \frac{H_{2}}{\hbar }-\omega
_{MW}+\Delta _{n}\right) \tau \right) \right| n\right\rangle =\left|
\left\langle n\left( t=0\right) \mid n\left( t=\tau \right) \right\rangle
\right| $. Since $\left\langle n\left| \exp \left( -i\left( \frac{H_{2}}{%
\hbar }-\omega _{MW}+\Delta _{n}\right) \tau \right) \right|
n\right\rangle =\left| \left\langle n\left| \exp \left( i\left(
\frac{H_{1}}{\hbar }\right) \tau \right) \exp \left( -i\left(
\frac{H_{2}}{\hbar }\right) \tau \right) \right| n\right\rangle
\right| $ this contrast serves as a measure of the stability of
the quantum evolution under a small perturbation in the
Hamiltonian \cite {Peres84}.

In our experiments atoms are not initially in a single motional
eigenstate, but a thermal ensemble of atoms incoherently populates
many eigenstates. The total population in $\left| 2\right\rangle $
is then given by an average of $P_{2}$ over the initial thermal
ensemble. $\Delta _{n}$ depends on the initial state, and due to
this spread in detuning the fringe contrast of the
ensemble-averaged $P_{2}$ decays rapidly even when $\left|
\left\langle n\left( t=0\right) \mid n\left( t=\tau \right)
\right\rangle \right| \simeq 1$ for all populated states (note the
analogy to the inhomogeneous broadening of two level atoms).
Hence, the decay of the fringe contrast in Ramsey spectroscopy of
thermal trapped atoms does not teach us anything about the
dynamics in the trap ($\left| \left\langle n\left( t=0\right) \mid
n\left( t=\tau \right) \right\rangle \right| $) but simply
measures the spread in $\Delta _{n}$. This spread also limits the
interrogation time for high precision spectroscopy.

Equation \ref{hami} indicates that if a special trap can be
constructed so that $\Delta _{n}$ is independent of $n$, then the
fast decay of the Ramsey fringes will be suppressed. An example of
such a dephasing-free trap is an inverted pyramid of exponentially
decaying evanescent waves, in which gravity provides the vertical
confinement. This trap have the special feature that $V_{2}\left(
{\bf x}\right) $ is exactly equal to $V_{1}\left( {\bf x}\right) $
shifted along the vertical axis. Therefore all eigenenergies are
shifted the same amount and $\Delta _{n}$ is indeed a fixed
number. This makes such a trap an excellent candidate for high
precision spectroscopy. Note, that in order to use traps for this
kind of measurements, the suppression of the relative spread in
$\Delta _{n}$, should be accompanied by a reduction in their
average value, e.g. by increasing the trap's detuning. Since the
proposed trap is a "dark optical trap", where atoms are trapped by
repulsive light forces mainly in the dark \cite{Friedman02}, hence
minimizing the amount of interaction with the trap's light, then
in addition to being dephasing-free it has also a low photon
scattering rate (which will now be the limiting factor for the
coherence time) and a small $\Delta _{n}$.

Apart from using special geometries a possible solution to the
above mentioned dephasing is to reverse it using "echo spectroscopy" by adding a MW $\pi $-pulse between the two $%
\frac{\pi }{2}$ pulses. $P_{2}$ for atoms initially populating a single
eigenstate then becomes:
\begin{eqnarray}
P_{2} &=&\frac{1}{2}\left[ 1-%
%TCIMACRO{\func{Re}}%
%BeginExpansion
\mathop{\rm Re}%
%EndExpansion
\left( \left\langle n\left| e^{i\left( \frac{H_{1}}{\hbar }\right) \tau
}e^{i\left( \frac{H_{2}}{\hbar }\right) \tau }e^{-i\left( \frac{H_{1}}{\hbar
}\right) \tau }e^{-i\left( \frac{H_{2}}{\hbar }\right) \tau }\right|
n\right\rangle \right) \right]  \nonumber \\
&=&\frac{1}{2}\left[ 1-%
%TCIMACRO{\func{Re}}%
%BeginExpansion
\mathop{\rm Re}%
%EndExpansion
\left( e^{i\frac{E_{n}^{1}}{\hbar }\tau }\left\langle \varphi _{n}\left(
t=0\right) \mid \varphi _{n}\left( t=\tau \right) \right\rangle \right) %
\right]  \label{formel1}
\end{eqnarray}
where $\left\{ \varphi _{n}\right\} $ is a new basis defined by $\left|
\varphi _{n}\left( t=0\right) \right\rangle \equiv \exp \left( -i\frac{H_{2}%
}{\hbar }\tau \right) $ $\left| n\right\rangle $, and $\left| \varphi
_{n}\left( t=\tau \right) \right\rangle \equiv \exp \left( -i\frac{H_{1}}{%
\hbar }\tau \right) $ $\left| \varphi _{n}\left( t=0\right)
\right\rangle $. $P_{2}$ no longer depends on $\omega _{MW}$ and
$E_{HF}$ and therefore not on any $\Delta _{n}$ but only on the
dynamics\footnote{$%
%TCIMACRO{\func{Re}}%
%BeginExpansion
\mathop{\rm Re}%
%EndExpansion
\left( \exp \left( i\frac{E_{n}^{1}}{\hbar }\tau \right)
\left\langle \varphi _{n}\left( t=0\right) \mid \varphi _{n}\left(
t=\tau \right) \right\rangle \right) \simeq \left| \left\langle
\varphi _{n}\left( t=0\right) \mid \varphi _{n}\left( t=\tau
\right) \right\rangle \right| $ is often a good approximation
\cite{Andersen02b}, but the limitations of it is not yet known to
the authors.}. Ensemble averaging over a thermal ensemble
therefore does not remove the fingerprints of the dynamics.

\begin{figure}[t]
\begin{center}
\includegraphics[width=3in]{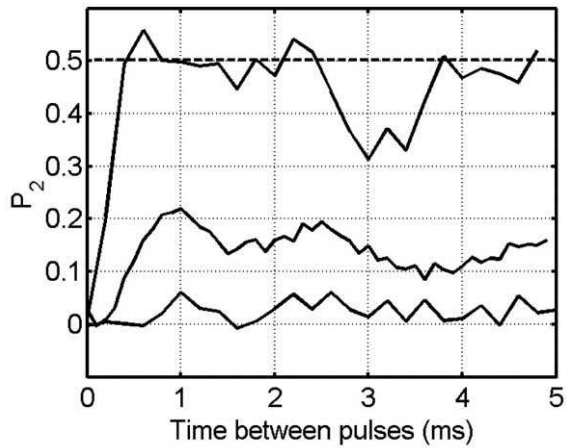}
\end{center}
\caption{Echo signal $P_{2}$ measured for three different trap
laser wavelengths $\protect\lambda $ close to the $D_{1}$
transition wavelength of 795 nm. Lower curve: A "small
perturbation" ($\protect\lambda $=805 nm). A good echo signal
($P_{2}<<1/2$) is seen almost independent of time between pulses.
Middle curve: A "medium perturbation" ($\protect\lambda $=798.25
nm). The echo signal
oscillates around a value smaller than 1/2 with partial revivals at $\protect%
\tau=1/2\protect\tau_{osc}$ and $\protect\tau=\protect\tau_{osc}$, where $%
\protect\tau_{osc}$=3.6 ms is the measured trap oscillation
frequency in the transverse direction. Upper curve: A "strong
perturbation" ($\protect\lambda $=796.25 nm). After a short time
the echo signal completely disappears ($P_{2}=1/2$), but partly
revives again at $\protect\tau=3.3$ ms, close to
$\protect\tau_{osc}$.} \label{fi2}
\end{figure}

We performed echo spectroscopy on $\sim 10^{5}$ $^{85}Rb$ atoms
laser cooled to $\sim 20$ $\mu K$ and then trapped in a $\sim 30$
$\mu K$ deep Gaussian optical trap with waist size of $\sim 50$
$\mu m$. The procedure is detailed in \cite{Andersen02b}. The echo
signal $P_{2}$ was measured as a function of time between pulses
for different wavelengths of the trap laser, thereby changing the
perturbation strength $\delta V$, while keeping the trap depth
constant by adjusting the power of the trap beam. For a small
perturbation a good echo ($P_{2}<<1/2$) is seen independent of the
time between pulses (see Fig. \ref{fi2}). For larger perturbation
damped oscillations to a level smaller than 1/2 are seen. An even
larger perturbation causes a complete decay of the echo coherence
($P_{2}=1/2$) followed by partial revivals, which are clear
signatures of the dynamics. Two revivals (of different amplitude)
are seen when the time between MW pulses is close to
$\frac{1}{2}T_{osc}$ and $T_{osc}$, respectively, where
$T_{osc}=3.5$ ms is the measured radial oscillation frequency of
the trap (the longitudinal oscillation frequency is $\sim 1$ $s$,
hence the longitudinal motion is essentially fixed for the
duration of our experiment). The revival at $\frac{1}{2}T_{osc}$
corresponds to wavepacket oscillations excited by the symmetric
change of the trap strength, and the stronger revival at $T_{osc}$
corresponds also to the vertical sloshing of wavepackets induced
by the vertical shift of the gravito-optical potential for $\left|
2\right\rangle $ compared to $\left| 1\right\rangle $ (in our
experiments this vertical shift is of the order of a few $nm$, as
compared to a trap cloud size of $\sim 50$ $\mu m$!). The lack of
complete revival at $T_{osc}$ is mainly due to anharmonicity of
the Gaussian trap, but also small amount of decoherence due to
spontaneous scattering of photons and noise may contribute.

The sensitivity of our technique to map the quantum dynamics of a
system is revealed here, since we see the dynamics due to a
perturbation (kick), that is about three orders of magnitude
smaller than $k_{B}T$.

\section{Eigenstate Stability and Atom Optic Billiards}
\begin{figure}[t]
\begin{center}
\includegraphics[width=3in]{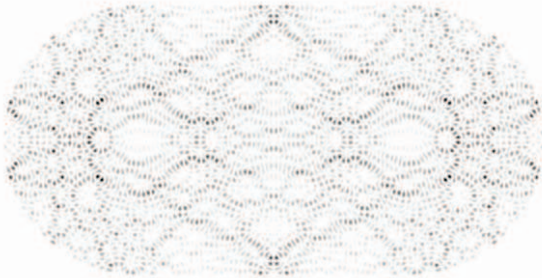}
\end{center}
\caption{Probability density plot of an eigenstate of a Bunimovich stadium (%
$k_{0}\simeq 100$).} \label{fi4}
\end{figure}

In Fig. \ref{fi2} it is seen that for sufficient large detuning
the long time asymptotic behavior of the echo signal from thermal
atoms is smaller than 1/2, and eventually becoming $\simeq 0$.
This can be understood from Eq. \ref{formel1}: If $\delta V$ is
small there will be, for each initial motional eigenstate $\left|
n\right\rangle $ of $H_{1}$\, a large matrix
element for going into the equivalent motional eigenstate of $H_{2}$, since $%
H_{1}$ and $H_{2}$ are almost identical. If the time between
pulses is large enough so the dynamics has dephased completely
between them, then the contributions from other states will
average to zero. Hence, the long time ensemble average of $P_{2}$
is simply given by $\overline{P_{2}} =\frac{1}{2}[ 1-
\overline{|\langle n^{\prime }=n|n\rangle |^{4}}] $, where $\left|
n\right\rangle $ and $\left| n^{\prime }=n\right\rangle $ are
corresponding eigenstates of $H_{1}$ and $H_{2}$ respectively. In
\cite{Andersen02b} a paraxial numerical calculation of
$\overline{P_{2}}$, averaged over the $\sim 4\times10^{6}$
thermally-populated states of the trap yielded good quantitative
agreement with measured long time echo signal.

\begin{figure}[t]
\begin{center}
\includegraphics[width=3in]{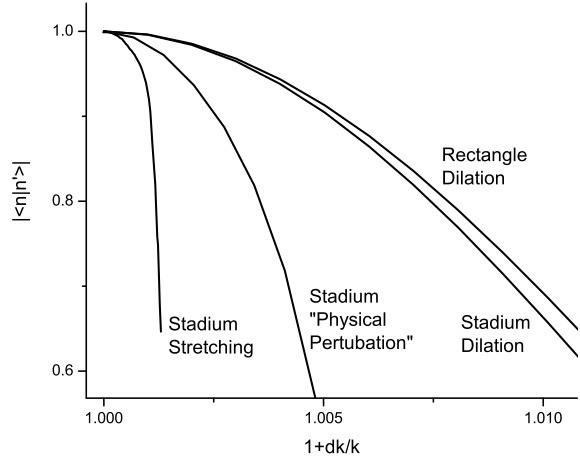}
\end{center}
\caption{$|\left\langle n^{\prime}=n|n\right\rangle |$ as a
function of perturbation strength $1+\delta k/k$ for eigenstates
with $k_{0} \simeq 100$. Three different perturbations are shown
for the stadium eigenstate of Fig. 3, namely dilation, stretching
and the "physical perturbation" (a dilation followed by a negative
stretch). The rectangle state has $k_{x} \simeq k_{y}$. For
dilation substantial reduction of $|\left\langle
n^{\prime}=n|n\right\rangle |$ indeed occurs for $\delta k \simeq
1$, as expected from the intuitive discussion in the text. }
\label{fi3}
\end{figure}

Because of the importance of the eigenstate stability to the
revival of coherence (good echo signal) and dynamics in the trap,
we now calculate $|\left\langle n^{\prime }=n|n\right\rangle |$
for perturbations, that we assume approximate our inherent
perturbation, i.e. the difference in dipole potential for the two
hyperfine states, for different traps. Note, that since a large
$|\left\langle n^{\prime }=n|n\right\rangle |$ also implies a
large survival probability as defined in
\cite{Peres84,Wisniacki01,Wisniacki02}, it also plays an important
role in quantum dynamical studies.

Atom optic billiards are good candidates for trapping atoms for
high precision spectroscopy, since the interaction with the trap
light is small because atoms are confined mainly in the dark
\cite{Friedman02}. They have been used to study dynamics in the
past, and have experimentally been demonstrated to display the
features of ideal billiards,
\cite{Milner01,Friedman01,Kaplan01,Andersen02}. For our
calculations we therefore approximate the atom optics billiard
with an ideal hard-wall billiard, and use the powerful technique
of Vergini and Saraceno to numerically calculate highly excited
eigenstates and eigenenergies \cite{Vergini95}. For an unperturbed
Bunimovich stadium \cite{Bunimovich79} (two semicircles of radius
$r=1$ connected by two straight lines of length $l=2$, for units
defined such that $\hbar=2m=1$) a plot of the probability density
of a highly excited state is shown in Fig.
\ref{fi4}.  The size of its wave vector is $k_{0}\simeq 100$ ($%
\nabla ^{2}\psi =$ -$k_{0}^{2}\psi $) hence the wavefunction have
on the order of few thousand antinodes.

\begin{figure}[t]
\begin{center}
\includegraphics[width=4.5in]{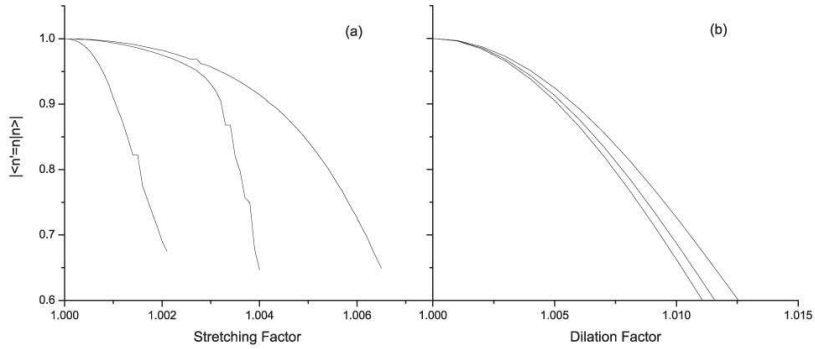}
\end{center}
\caption{$|\left\langle n^{\prime}=n|n\right\rangle |$ for three
states of the stadium with $k_{0} \simeq 100$ as a function of
perturbation strength. (a) Stretching: A strong dependence on the
initial state is seen. (b) Dilation, for the same three stases as
in (a): Only a weak dependence on the initial state with similar
$k$ is seen.} \label{fi5}
\end{figure}

Systems respond very different to different types of perturbations
\cite{Barnett00,Cohen00,Cohen01,Wisniacki01,Wisniacki02}.
Perturbations of $V_{1}\left( {\bf x}\right) $ that can be
described as a change of coordinate system such as rotation,
translation (of importance to the inverted pyramid discussed
above) and dilation will give exceptional high $|\left\langle
n^{\prime }=n|n\right\rangle |$. Intuitively this is clear since
an eigenstate of $H_{1}$ under these three perturbations is also
an eigenstate of $H_{2}$ rotated, translated or
dilated respectively. Therefore, to cause a significant drop in $%
|\left\langle n^{\prime }=n|n\right\rangle |$ the perturbation has
to move the billiard wall distances on the order of typical
distances between nodes in the eigenstate wave function,
corresponding to $\delta k\simeq 1$ in case of dilation (see Fig.
\ref{fi3}). On the other hand perturbations that changes the
symmetry of the billiard can cause much faster changes in
$|\left\langle n^{\prime }=n|n\right\rangle |$. This is
illustrated in Fig. \ref{fi3} where it is seen that dilating a
Bunimovich stadium causes $|\left\langle n^{\prime
}=n|n\right\rangle |$ to drop much slower than changing the
lengths of the straight sections (this perturbation we call
"stretching"). We are interested in the perturbation induced in an
atom-optics billiard by changing the strength of the optical
potential. This is best approximated by moving the billiard's wall
a fixed distance perpendicular to itself. This perturbation, that
we name ''the physical perturbation'', can be viewed as a dilation
followed by a negative stretch, and it shows features between
those of dilation and stretch (Fig. \ref{fi3}).

\begin{figure}[t]
\begin{center}
\includegraphics[width=3in]{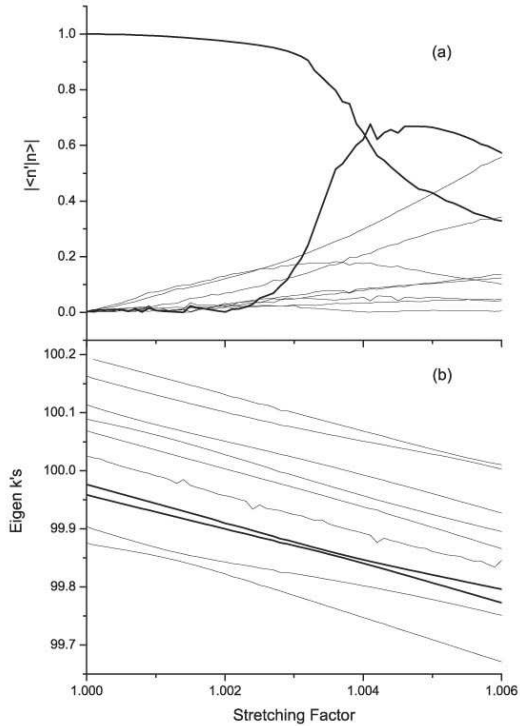}
\end{center}
\caption{(a) Matrix elements ($|\left\langle
n^{\prime}|n\right\rangle |$) between an initial state and its ten
closest neighboring states with the same parity, as a function of
stretching factor. (b) Eigen $k$'s as a function of stretching
factor, for the same states as (a). It is seen that when the state
undergo avoided crossing it mixes strongly with the crossed state.
Bold lines in (a) and (b) corresponds to the same states. }
\label{fi6}
\end{figure}

In Fig \ref {fi3} $|\left\langle n^{\prime }=n|n\right\rangle |$
for a rectangle is also shown in order to illustrate that the
criteria stating that the dilation should move the wall of the
billiard a distance of the order of the distance between nodes is
valid not only for chaotic shapes but also for regular ones. For
dilation in 1D systems this criteria is equivalent to the
necessary criteria for substantial mixing from perturbation
theory, namely that the perturbation energy should be comparable
to the level separation. That this criteria also holds for
dilating an integrable shape in 2D (where it corresponds to a
perturbation energy of $\sim 100$ times the level spacing in our
case) is also obvious since it is equivalent to two 1D systems.
But its fulfilment for a chaotic system where variables cannot be
separated is less obvious unless the above arguments are
considered.

\begin{figure}[t]
\begin{center}
\includegraphics[width=3in]{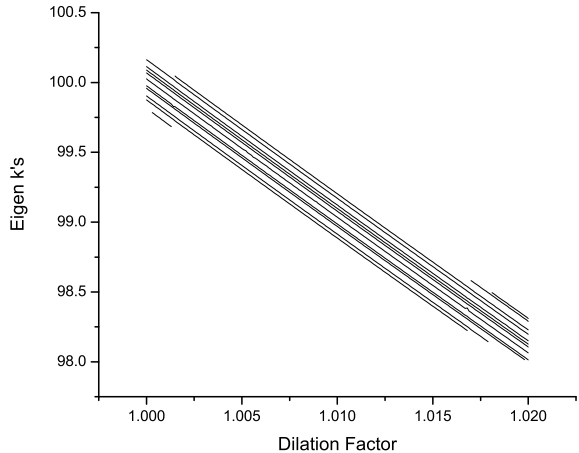}
\end{center}
\caption{$k$ as a function of dilation factor for eigenstates with
$k_{0} \simeq 100$ in a stadium. Straight lines are observed as
excepted.} \label{fi7}
\end{figure}

Figure \ref{fi5} shows $|\left\langle n^{\prime }=n|n\right\rangle
|$ as a function of perturbation strength for different
eigenstates with $k_{0}\simeq 100$. It is seen that $|\left\langle
n^{\prime }=n|n\right\rangle |$ is highly state dependent in case
of stretching, but is not in case of dilation. Following the
explanation given above, it is clear that states with almost the
same $k_{0}$ have almost the same $|\left\langle n^{\prime
}=n|n\right\rangle |$ for a given perturbation strength in case of
dilation. The sudden drop in $|\left\langle n^{\prime
}=n|n\right\rangle |$ in case of stretching is often associated
with avoided crossings of energy levels, after which the two
states involved still hold the major part of the probability. This
is clearly seen from Fig. \ref{fi6}, where matrix elements to the
ten closest states of same parity are shown for
stretching\footnote{Stretching is an even parity perturbation and
selection rules therefore insures coupling only to states with the
same parity}. In Fig. \ref{fi6}b $k$ as a function of stretching
factor is shown for the same ten states. As seen, the avoided
crossing of the two neighboring states (drawn in bold lines)
occurs for the same stretch factor as the one causing strong
mixing between them (Fig. \ref{fi6})). In Fig. \ref{fi7} $k$ as a
function of the dilation factor is presented, showing straight
lines as expected. Matrix elements to close by states are 0 for
dilation, in strong contrast to stretching shown in Fig.
\ref{fi6}a \cite{Barnett00}.

\section{Conclusions}

We have analyzed echo spectroscopy of trapped ultra-cold atoms.
Using this technique, we can combine the high precision of
MW-spectroscopy with the high degree of control over experimental
parameters provided by laser cooling and trapping techniques, to
perform quantum dynamical studies even when many states are
thermally populated. Based on this analysis we proposed a
gravito-optical trap in which atoms in superposition states does
not suffer from dephasing due to inhomogeneous Stark shifts.

We analyzed numerically the stability of eigenstates for several
perturbations, and verified that the rapid drop in $|\left\langle
n^{\prime }=n|n\right\rangle |$ seen for symmetry breaking
perturbations is linked to avoided crossings of energy levels. The
symmetry of the effective perturbations, caused by MW transitions
in our experimental atom optics billiards was shown to depend on
their shape. For example, rectangle or a circular billiard the
physical perturbation corresponds approximately to pure dilation,
thus enhancing the applicability of echo spectroscopy for these
shapes, as compared to the Bunimovich stadium billiard discussed
above.

Finally, the suppression of dephasing achieved with echo
spectroscopy yields a dramatic increase in the coherence time for
trapped atoms, that may find important applications for precision
spectroscopy and quantum information processing.

\textbf{Acknowledgements}

This work was in part supported by Minerva Foundation, the United
States-Israel Binational Science Foundation, and Foundation
Antorchas. M. F. A. Acknowledges help from Dansk-Israelsk
Studiefond til Minde om Josef og Regine Nachemsohn.

\end{document}